\documentclass[10pt,journal]{IEEEtran}

\usepackage[pdftex]{graphicx}
\DeclareGraphicsExtensions{.pdf,.jpeg,.png}

\usepackage{multirow,setspace,verbatim,amsfonts,graphicx,amsmath,amsthm,amsbsy,amssymb,epsfig}
\usepackage{url,cite}
\usepackage{booktabs}

\usepackage{array}
\usepackage{booktabs}

\begin{document}

\title{GSSMD: New metric for robust and interpretable assay quality assessment and hit selection}
\author{Seongyong Park, Shujaat~Khan
	\thanks{The authors are with the Department of Bio and Brain Engineering, Korea Advanced Institute of Science and Technology (KAIST), Daejeon, Republic of Korea. Email:\{sypark0215, shujaat\}@kaist.ac.kr}
	\thanks{The implementation of this methods in both MATLAB and R are available at the author's github page (https://github.com/psychemistz/gssmd). We also provide scripts to reproduce all figures in our manuscript with R and MATLAB.}}

\maketitle

\begin{abstract}
In the high-throughput screening (HTS) campaigns, the Z'-factor and strictly standardized mean difference (SSMD) are commonly used to assess the quality of assays and to select hits. However, these measures are vulnerable to outliers and their performances are highly sensitive to background distributions. Here, we propose an alternative measure for assay quality assessment and hit selection. The proposed method is a non-parametric generalized variant of SSMD (GSSMD). In this paper, we have shown that the proposed method provides more robust and intuitive way of assay quality assessment and hit selection.
\end{abstract}


\IEEEpeerreviewmaketitle

\section{Introduction}
In high-throughput screening, hundreds to tens of millions\cite{broach1996high} single measurements of test samples are often compared to positive and negative control to characterize the effectiveness of the treatment. For instance, in a typical drug screening, luminescence-based cell viability assays can be used to test the cytotoxic effects of compounds in comparison to known drugs and biological inerts. Since the performance of the assay is assessed retrospectively, the assay developer must measure the identifiability and reproducibility of the assay to determine the appropriate one from the candidates before conducting the actual HTS campaign\cite{Malo2006StatisticalPI}. Within an acceptable assay, assay developer also need to set the threshold of acceptance according to the quality of the assay. 

Quality metrics such as Z'-factor\cite{Zhang1999ASS} and strictly standardized mean difference (SSMD) \cite{Zhang2007APO} have been used for this purpose, but these quality measures have several limitations, e.g., sensitivity to measurement noise etc. Only statistical parameters such as the mean and variance of the distribution are taken into account, which makes them vulnerable to outliers in the measurement. In addition, since infinite scale ranges of these measures are generally not intuitive for most biologists \cite{birmingham2009statistical}, it can be complicated to choose suitable thresholds for various types of experiments and it is not easy to understand the meaning of these values without referring to the literature. These disadvantages limit the applicability of these measures.

To address aforementioned issues, we propose a non-parametric generalized variant of strictly standardized mean difference (GSSMD). The proposed GSSMD is inspired by a recently proposed contrast measure\cite{rodriguez2019generalized} for ultrasound imaging systems. GSSMD is defined using the overlap between two distributions. Thus, it avoids the complexity of Z'-factor and SSMD calculations in non-standard or transformed distributions. By definition, it is intuitive enough and easy to interpret. In this paper, we used simulations and real biological data sets to support our claims. Mathematical definitions and experimental results are described in the following sections.

\section{Background}\label{background}
\subsection{Conventional measure of assay quality}
Classic indicators of assay quality are the signal-to-noise ratio (SNR) and the signal-to-background ratio (SBR)\cite{Malo2006StatisticalPI}. The SNR and SBR are defined as $S/N = (\mu_1-\mu_2)/\sigma_2$ and $S/B = \mu_1/\mu_2$ where $\mu$ and $\sigma$ are the mean and variance of the distribution. The more recently proposed metric is called Z'-factor\cite{Zhang1999ASS} and has been proposed to provide a better metric considering the dynamic range of measurement. The Z'-factor is defined as:
\begin{equation}\label{z-factor-equation}
Z'_f = 1-\frac{3(\sigma_1 + \sigma_2)}{|\mu_1 - \mu_2|}    
\end{equation}
One major drawback of the Z'-factor comes from the normal distribution assumption behind it. Because of this assumption, it is often too strict for many biological assays that are fully functional but do not follow normal distribution. Moreover, Z'-factor is difficult to interpret due to the imbalanced range given by $-\infty \leq Z'_f \leq 1$.

A slightly improved metric for assay quality is SSMD\cite{Zhang2007APO} which has been proposed to solve the problem of Z'-factor mentioned above. The SSMD is defined as:
\begin{equation}\label{ssmd-equation}
\text{SSMD} = \frac{\mu_1 - \mu_2}{\sqrt{\sigma_1^2 + \sigma_2^2}}
\end{equation}
Although SSMD resolves some of the issues of Z'-factor such as unbalanced range and too strict thresholds but is still sensitive to the types of distribution and distribution transformations. Also, similar to Z'-factor, an infinite range of SSMD ($-\infty \leq \text{SSMD} \leq \infty$) makes it difficult to set the correct criteria for each assay.

\subsection{Contrast-to-noise ratio and its relationship with assay quality measures}
The quality of assay can be defined as the contrast between two sets of measurements. If two sets of measurements are significantly different from each other, we can say that the given measurements are obtained from high quality assay. Interestingly, this contrast-based quality assessment is widely used for medical imaging, and similar metrics are proposed for relatively similar tasks. For example, in the field of ultrasound imaging, the contrast of intensity in two regions of interest is used for the identification of anatomical structures and associated abnormalities or diseases. A widely used measure of contrast is contrast-to-noise ratio (CNR) \cite{rodriguez2019generalized}. The formula of CNR is shown below.

\begin{equation}\label{cnr-equation}
\text{CNR} =  |\text{SSMD}| = \frac{|\mu_1 - \mu_2|}{\sqrt{\sigma_1^2 + \sigma_2^2}}
\end{equation}
where $\mu$ and $\sigma$ are the mean and variance of the intensities in two regions.

From the above equations \eqref{ssmd-equation} and \eqref{cnr-equation}, it is easy to see that the CNR measure is very similar to the SSMD. The only difference is that the CNR is suitable magnitude of the change only, while SSMD considers both the direction and the magnitude of the change. 

The CNR has two major limitations. First, it is sensitive to distribution transformations. Therefore, a high CNR values does not mean high detectability especially if underlying distributions are different. Second, it is not intuitive because the range of scale is infinite\cite{rodriguez2019generalized}. These are the same limitations as SSMD we mentioned earlier. 

\subsection{Proposed generalized variant of SSMD (GSSMD)}
Recently, to address limitations of CNR, a generalized variant of CNR called GCNR is proposed \cite{rodriguez2019generalized}. It is defined as:
\begin{equation}\label{gcnr-ol-equation}
\text{OVL} = \int \min\{p_1(x), p_2(x)\} dx    
\end{equation}
$$\text{GCNR} = 1-\text{OVL}$$
where OVL is the overlap of two probability density functions (PDFs) $p_1(x)$ and $p_2(x)$. The range of overlap is $0 \leq \text{OVL} \leq 1$.

GCNR is intuitively defined as non-overlapping proportion between two PDFs. In addition, unlike CNR or SSMD, GCNR is less susceptible to distribution transforms because there is no underlying assumption about the distribution. In practice, the PDFs of positive and negative control samples can be estimated non-parametrically by histogram with an appropriate bin size, depending on the number of samples \cite{cellucci2005statistical}.

Inspired by GCNR, here we propose generalized variant of SSMD (GSSMD) to address similar problems in the field of assay development. As in the case of SSMD and CNR, the proposed GSSMD is similar to GCNR, but unlike GCNR where only the size of the overlap area is taken into account, GSSMD also considers the direction of overlap. Mathematically the GSSMD is defined by the formula shown below.

\begin{equation}\label{gssmd-equation}
\text{GSSMD} = \text{sgn}(\mu_1 - \mu_2)\times(1-\text{OVL})    
\end{equation}
where $\text{sgn}[.]$ is sign operator and the GSSMD is in the range $-1 \leq \text{GSSMD} \leq 1$.

The following sections describe the performance and effectiveness of the proposed method using simulations and real biological experiment data.

\section{Results}
In this study, Z'-factor, SSMD and GSSMD are compared in simulation and real biological experimental data set. For the simulation setup, different configurations are selected for the distribution of positive and negative samples, each of which mimics a specific case of study. For experimental data set, we considered RNAi screening of cell viability in Drosophila Kc167 cells \cite{boutros2004genome} as a case study. For all experiments, to estimate the histogram, we used $1+log2(N)$ number of bins\cite{cellucci2005statistical}, where $N$ is the number of samples.

\subsection{Simulations}
In our hypothetical experimental setup, we compared performance and interpretability of the Z'-factor and SSMD with proposed GSSMD in different cases. We studied the relationship between the mean difference of the two distributions and the calculated value of each measure. We focused on normal distributions with equal variances and then we looked at the log-normal distributions that are often observed in biological systems\cite{furusawa2005ubiquity}. We also identified the effect of outliers on measurements derived from known mean and variance of the normal distribution. To account for various types of measurement noise, we investigated the effect of additive white Gaussian noise on sampled data. We compared SSMD and GSSMD at different noise levels controlled by the signal-to-noise-ratio ($\text{SNR}_{measurement}$) of the measurement. Finally, to set the lower bound of GSSMD from the variances associated with non-parametric estimation of PDFs, we examined the effect of sample number on GSSMD calculations.

\subsubsection{Normal distribution with equal variance}

\begin{figure*}[htbp]
\centerline{\includegraphics[width = 15cm]{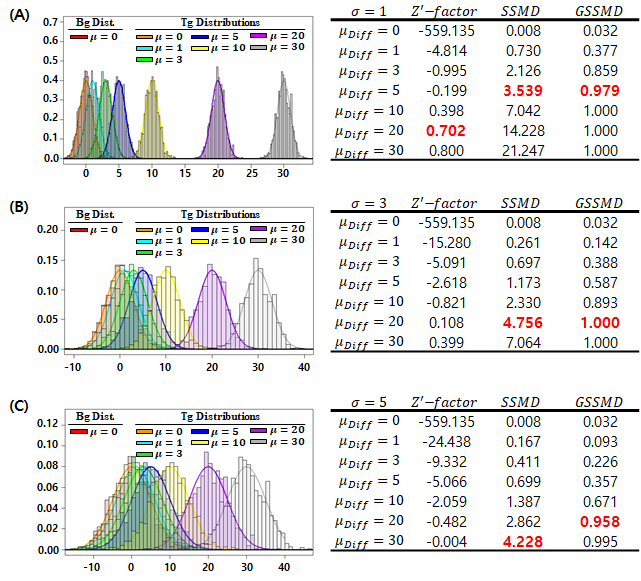}}
\caption{GSSMD provided interpretable and reliable statistical difference estimate in the ideal simulation setup ($\sigma_{c}=1, 3, 5$ , $n = 1,000$ in (A), (B), and (C) respectively). The numbers in red in the table represent values that have exceeded acceptable quality thresholds for the assay, according to the Zhang\cite{Zhang1999ASS, Zhang2007APO}, excluding GSSMD. Here we set the threshold for GSSMD to $\text{GSSMD}=0.95$ which represents a $5\%$ overlap between the two distributions. Bg Dist. = Background Distribution, Tg Distributions = Target Distributions.}
\label{fig1}
\end{figure*}

In this experiment, three cases were selected for the distribution of negative control samples. In particular, for the distribution of negative control samples the mean is assumed to be zero $\mu=0$ in all three cases, but the variance changes to $\sigma=1, 3$ and $5$ (see Figure\ref{fig1}). We generated $21$ positive control sample distributions drawn from $\mathcal{N}(\mu, \sigma)$ where $\mu=0, 1, 3, 5, 10, 20$, and $30$ and $\sigma=1, 3$ and $5$. Figure~\ref{fig1}., show the distribution and histogram of each case for each measure. 
In the Figure~\ref{fig1}., GSSMD increases as the mean difference increases and it decreases as the variance of the distribution increases. Note that in Figure~\ref{fig1}.(A), GSSMD is maximum (i.e., equals to one) with mean difference of $10$ (yellow curve), so there is no overlap between the two distributions. In other cases (see Figure \ref{fig1}(B and C)) due to an increase in distribution variance, the value gradually decreases, indicating an increase in overlap. On the other hand, the values of Z'-factor and SSMD change dramatically and do not mean anything by themselves.
In addition, GSSMD exhibits the highest sensitivity in terms of acceptable levels of assay quality, which is set at $0.95$ to indicate $5\%$ overlap between the positive and negative control distributions. The acceptable quality scores of Z'-factor and SSMD are 0.5 and 3, referenced in Zhang\cite{Zhang1999ASS, Zhang2007APO}.
Although Z'-factor or SSMD can be applied to measure effect size of large differences, in many real data sets there is no clear separation, as shown in the Figure~\ref{fig1}.(A), so GSSMD is practically better suited as a quality assessment method. This is further evaluated in later sections. 

\subsubsection{Log-normal distribution with equal variance}

\begin{figure*}[htbp]
\centerline{\includegraphics[width = 15cm]{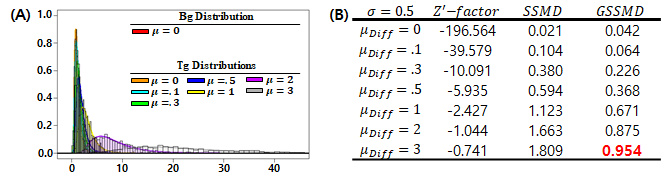}}
\caption{Performance of each measure in log-normal distribution setup ($\sigma=0.5$). The acceptable values are $\geqslant0.5$, $\geqslant3$, and $\geqslant0.95$ respectively.}
\label{fig2}
\end{figure*}

We also compared each measure with the non-normal distribution setting. For this purpose, log-normal distribution was chosen because the data obtained from biological experiments often follow the log-normal distribution \cite{furusawa2005ubiquity}. The recommended acceptable thresholds for Z'-factor, SSMD and GSSMD are $0.5$ \cite{Zhang1999ASS}, $3$\cite{Zhang2007APO}, and  $0.95$ respectively. As you can see in the Figure~\ref{fig2}., GSSMD can effectively detect the difference between the two distributions when mean difference is $\mu_{diff} = \mu_1 - \mu_2 = 3.0$, and variance is $\sigma = 0.5$, contrary other measures did not detect any difference in acceptable assay quality for any given configuration. 

\subsubsection{Effect of outliers}
Unlike simulation, real experimental measurements are prone to noise and outliers. To simulate the effect of outliers, we replaced some positive samples in the original sample set with outlier samples which are obtained from an independent normal distribution. The original sample set consist of $N_t = 2000$ samples, ($N=1,000$) in each class (positive and negative). Both the positive and negative samples are drawn from the identical normal distributions ($\mathcal{N}(0,1)$). Since two distributions are identical therefore the expected $\text{GSSMD} = 0$. 

\begin{figure*}[htbp]
\centerline{\includegraphics[width=18cm]{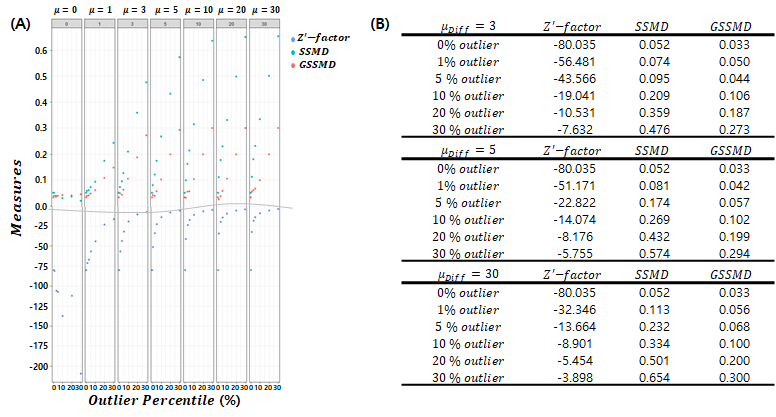}}
\caption{GSSMD is relatively insensitive to outliers compared to Z'-factor and SSMD. Blue, green and red dots represent Z'-factor, SSMD and GSSMD respectively.}
\label{fig3}
\end{figure*}

In particular in this experiment the outliers effect is simulated by replacing positive samples with outlier samples drawn from six different normal distributions. The outliers are generated using $\mathcal{N}(\mu,\sigma)$, where the variance $\sigma = 1$ for all distributions and the mean values are $\mu = 1, 3, 5, 10, 20$ and $30$. For each case, $1,000$ positive and negative samples were generated and $0\xrightarrow{}30\%$ positive samples were replaced with outlier samples.

Figure~\ref{fig3}. shows the effect of outliers on three measures, where the blue, green and red dots represent Z'-factor, SSMD and GSSMD respectively. In the subplot, we can see that when we increase the proportion and mean values of the outliers, the Z'-factor changes sharply and its absolute value become $\sim 18$ times smaller than outlier free condition. 

GSSMD, on the other hand, converges to the percentage of outliers. Especially when the mean of outlier samples is greater than $5$. In comparison to Z'-factor, SSMD measure is relatively insensitive to outliers and approaches to a specific value when the mean of outlier samples is greater than $10$. However, the values of SSMD has no intuitive meaning. Thus, GSSMD is the most sensitive measure for detecting outliers at small mean differences and the values have intuitive meaning, indicating the percentage of outliers. 

\subsubsection{Effect of signal-to-noise-ratio (SNR)}
There is other types of noise associated with measurement inconsistency. This type of noise is modeled as Gaussian noise ($\mathcal{N}$) with different level of SNR. Here the term SNR is different from assay quality metric mentioned earlier, which is given as $\text{SNR}_{assay}=(\mu_1-\mu_2)/\sigma_2$. The SNR of measurement is defined as the logarithm of the power ratio, $\text{SNR}_{measurement}=10 \times log_{10}(P_{signal}/P_{noise})$. We used MATLAB built-in function \textbf{\textit{awgn}} to add white Gaussian noise into our simulated measurement signal. 
Here the background signal is defined as a randomly sampled data points drawn from $\mathcal{N}(0,1)$ and the target signal is defined as \emph{background signal} + \emph{defined mean difference}. We added white Gaussian noise independently for each signal and calculate SSMD and GSSMD to clarify the difference between two measures in terms of measurement signal-to-noise ratio. 

\begin{figure*}[htbp]
\centerline{\includegraphics[width=15cm]{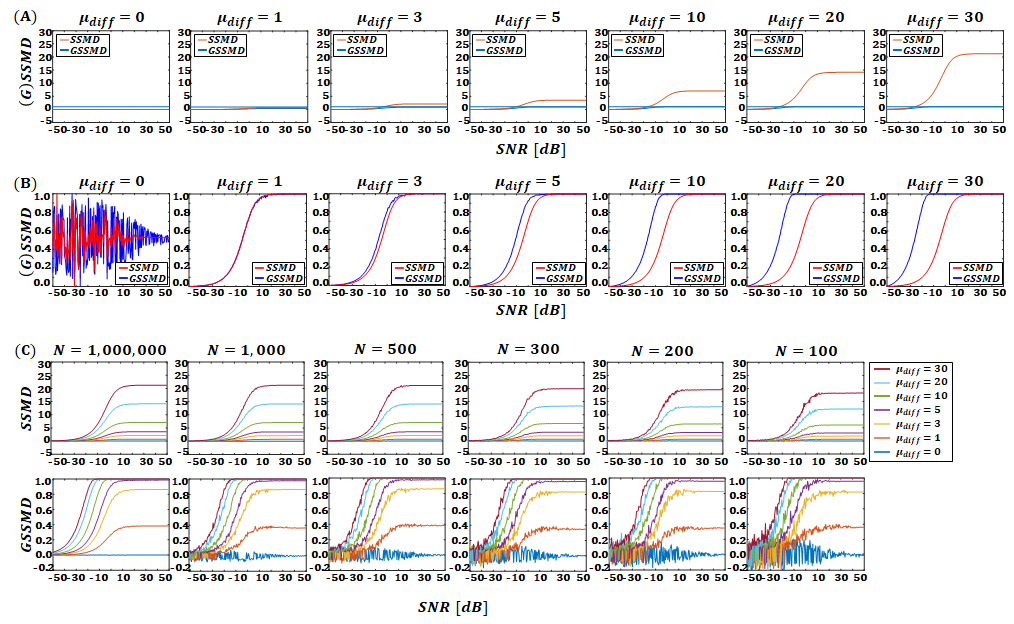}}
\caption{GSSMD is better quality metric for the noisy data sets.
(A) The range of SSMD increases with the mean difference, so the meaning of the values can vary depending on the mean difference. Unlike SSMD, GSSMD is limited from 0 to 1 (in case of positive overlap), so the meaning of the value is not affected by the mean difference.
(B) Scaled plot for each measure. GSSMD is more sensitive when the mean difference is large. Here we scaled the value from 0 to 1.
(C) SSMD and GSSMD with various sample sizes, noise levels and mean difference. GSSMD identifies mean differences even in the high noise regions.
x-axis represents SNR in deci-Bell (dB) scale for all figures.}
\label{fig4}
\end{figure*}

Figure~\ref{fig4}(A). show estimated SSMD and GSSMD at different noise levels and mean differences. Here the number of samples is quite large ($N=1,000,000$). As you can see, SSMD scale varies when the mean difference increases however GSSMD range doesn't changes. GSSMD represent non-overlap proportion by definition, it is always from -1 to 1 but we only consider positive mean difference in this example, so it changes from 0 to 1. Figure~\ref{fig4}.(B), shows scaled version of figure~\ref{fig4}.(A), to clarify the difference between SSMD and GSSMD. The blue line represents GSSMD and the red line represents SSMD. If there is no mean difference, compared to SSMD, the GSSMD due to the stochastic nature of sampling exhibits high fluctuation, especially in the region of low SNR. The value range for GSSMD is about $\pm 0.003$ but here we adjusted it from $0$ to $1$ to compare it with other conditions. However this fluctuation decreases with increasing SNR. If the mean difference between the two distributions is equal to one, GSSMD has the same sensitivity as SSMD. In other cases where the mean difference between the two distributions is larger, the sensitivity of GSSMD to SNR is greater than that of SSMD. 

One important aspect of quality metric is the sensitivity to the number of measurement samples. Here we investigated the effect of sample size as shown in the Figure~\ref{fig4}.(C). Each line in the subplot of Figure~\ref{fig4}.(C), represents different effect size ($\mu_{diff}=0, 1, 3, 5, 10, 20$ and $30$). The $x$ and $y$-axes of the plot represent the SNR in the deci-Bell (dB) scale and the measured SSMD and GSSMD values respectively. 

\begin{figure*}[htbp]
\centerline{\includegraphics[width = 15cm]{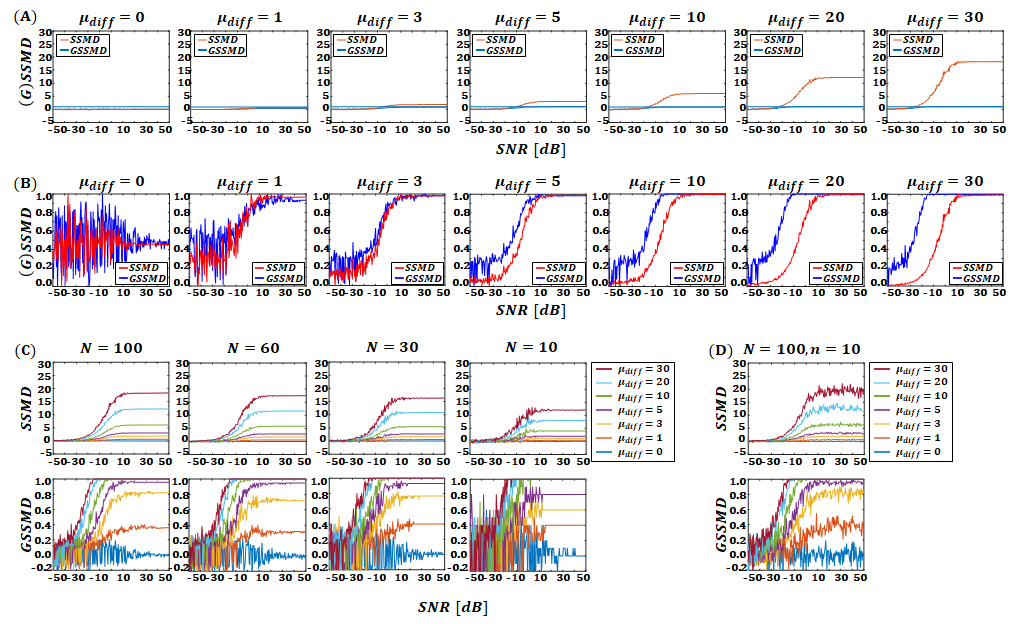}}
\caption{GSSMD can be used for small and noisy data sets. (N < 100)
(A) SSMD and GSSMD with sample size 100. The curve is little noisier but the trend is maintained. 
(B) Scaled plot for each measure. GSSMD is more sensitive when the mean difference is large, as shown in the figure 4 (B). Here, we adjusted the value from 0 to 1 too.
(C) SSMD and GSSMD with sample size less than 100. The signal is too deformed to distinguish mean difference when N = 10. 
(D) GSSMD and SSMD estimation using resampling. GSSMD and SSMD can be estimated by sampling 10 of the 100 samples. (10 runs) The sensitivity of GSSMD is slightly reduced but still better than SSMD.}
\label{fig5}
\end{figure*}

For further validation, the effect of smaller sample sizes was investigated. Figure~\ref{fig5}.(A), shows the trend of SSMD and GSSMD at different noise levels and the mean differences. Here we performed the same experiment as in Figure~\ref{fig4}.(A), and (B), with fewer number of samples ($N = 100$). Figure~\ref{fig5}.(B), show scaled version of Figure~\ref{fig5}.(A). As expected, the decrease in the number of samples caused the SSMD and GSSMD to fluctuate due to the uncertainty in the PDF estimation. However, the GSSMD has better resolution than SSMD even with this small sample number.

In Figure~\ref{fig5}.(C), we showed that the GSSMD can be applicable to even smaller number samples i.e., ($N \leq 100$). Although $N\leq10$ is very small for HTS setting however, in lab scale settings we often observe small samples size i.e., $\simeq10$ samples for one condition. The calculated GSSMD does not produce meaningful results in the low SNR region of $N\leq10$. This situation can be handled with the sub-sampling technique shown in Figure~\ref{fig5}.(D). Here, we combined $n=10$ independent batch experiments with $N=10$ samples each ($N_{total}=100$ total) and measure each metric based on $N=10$ samples ($n=10$) times of sub-sampling. We used $N=10$ data points to estimate SSMD and GSSMD for each sub-sampling and showed the average results. The results showed similar power as $1$ independent batch experiment with $100$ samples.

\subsubsection{Lower bound estimation for worst case scenario}
Since GSSMD is a non-parametric measure, the value of GSSMD may not always be zero if there is no mean difference between the two PDFs. This is true because the accuracy and stability of GSSMD metric depends on the number of samples used in the estimation of two PDFs. Therefore, we decided to investigate lower bound of GSSMD measure when there is a complete overlap between the two PDFs (the worst-case scenario). 

In this experiment, we estimated the lower-bound of GSSMD for two sample sets. In particular, we tested normal and log-normal distribution cases both with $\mu=0$ and $\sigma=1$. Theoretically, the expected GSSMD value for both scenarios is $0$, but some variation in the expected value might occur due to the variations in estimation of PDFs. Therefore, we performed an $10,000$ independent trials and calculated the mean, variance and extreme values of GSSMD for each sample size.

\begin{figure*}[htbp]
\centerline{\includegraphics[width=15cm]{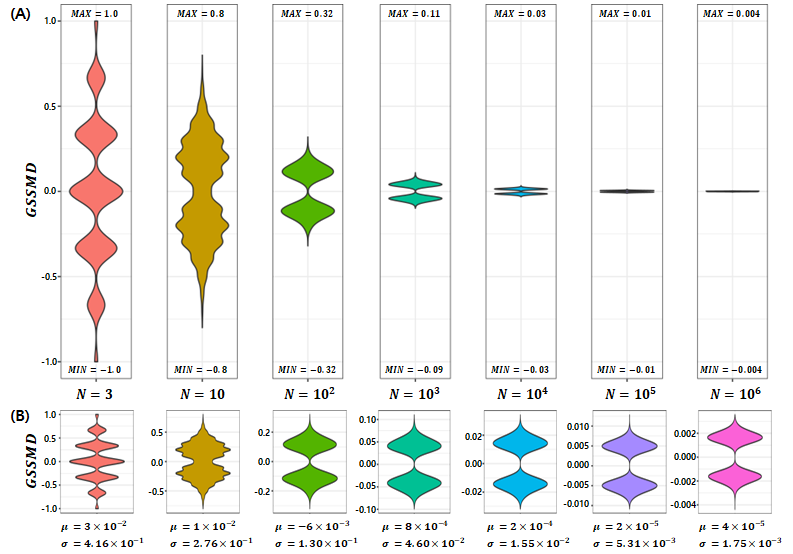}}
\caption{GSSMD score for overlapping distributions, effect of sample-size and lower bound for reliable screening. (A) Violin plot of GSSMD range -1 to 1, (B) Violin plot of GSSMD with matched scale.}
\label{fig6}
\end{figure*}

Figure~\ref{fig6}. show the estimated GSSMD for various sample size ($N=3\xrightarrow{}10^{6}$). As expected, as the sample size increase, the mean and variance of the estimated GSSMD decrease. We also found that the shape of distributions of estimated GSSMD with the sample size of more than $100$ is stabilized. The lower-bound of GSSMD for a specific sample size can be seen in the figure. Here we showed only normal distribution case, but the overall behavior is quite similar for both distributions (\textit{see}: https://github.com/psychemistz/gssmd). 

For log-normal distribution the lower-bound is even smaller than the normal distribution case, however for simplicity one can choose the maximum of the normal distribution to avoid possible false positives. This robust and intuitive nature of GSSMD measure make it a suitable choice for reliable assessment of assay quality. To reject this null hypothesis, throughout our simulation study, we chose $5\%$ as the threshold of GSSMD for sample size $N=1,000$.

\subsection{Case study of RNAi screening using real experiment data}
For the case study, we selected RNAi screening of cell viability in Drosophila Kc167 cells \cite{boutros2004genome}. We used the dataset provided by cellHTS2 R package\cite{boutros2006analysis}. In RNAi screening experiment HTS of $19,470$ dsRNAs was performed in Kc167 cells to characterize Drosophila genes, which are related to cell growth and viability. For every well of each plate, the authors introduced RNAi of different genes and measured cell viability through Luciferase activity. 

\begin{figure}[htbp]
\centerline{\includegraphics[width=9cm]{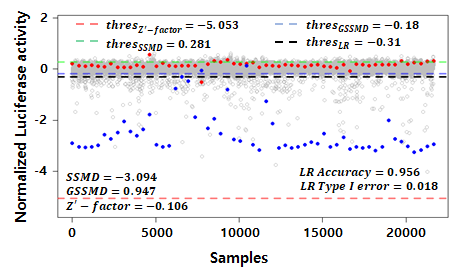}}
\caption{GSSMD can be used to find hit thresholds in RNAi screening. We set the GSSMD threshold to a point where the overlap of the two distributions is less than $5\%$. GSSMD threshold was similar to the threshold of optimal logistic regression classifier (Accuracy = 0.956, Type I error = 0.018) but the recommended threshold based on Z'-factor and SSMD were either too strict or too loose to identify the actual hit.}
\label{fig7}
\end{figure}

In other words, when targeted binding of specific RNAi promotes a decrease in cell viability, Luciferase activity was reduced according to the cell viability.  The Figure~\ref{fig7}. shows the calculated measures on the dataset, where red dots represent negative controls (GFP, Rho, no RNAi treatment) and blue dots represent positive controls (RNAi for D-IAP1 which inducing time-depedent cell death in Drosophila Kc167 cells). Grey dots are samples treated with RNAi, and it is not known how much reduction in cell viability indicates targeted RNAi binding.

Like Z'-factor and SSMD, the proposed GSSMD can also be used to identify thresholds for hit selection. In order to evaluate performance of each measure for finding the correct threshold of hit selection, we trained logistic regression classifier based on positive and negative control samples. The logistic regression class separation boundary is used a reference for optimal threshold of Luciferase activity. We trained the logistic regression classifier on the first plate of the replicated experiment, and tested it on the second plate. The threshold provided by the optimal logistic regression classifier (accuracy = 0.956, type I error = 0.018) was similar to the GSSMD threshold ($5 \%$ overlap), but the recommended thresholds based on the Z'-factor and SSMD were either too strict or too loose. Such strict or loose thresholds can result in the loss of many functional targets or in too many false positives, as clearly demonstrated in this example. 

\section{Discussion}
Bio-assay quality assessment and threshold setting for the hit selection are routinely performed in academia and industrial environments, where a large number of single measurements are performed. Recent advances in biotechnology have enabled new types of large scale experimental assays including but not limited to CRISPR-Cas9  \cite{ran2013genome} screening and cell painting \cite{bray2016cell}. However, current measures like Z'-factor and SSMD are designed to assess the quality of high-throughput chemicals and RNAi screening and rely heavily on normal distribution assumption. Therefore, it is not clear whether these measures can be applied to detect useful hits in newly developed assays with sufficient sensitivity. 

Also, as described in the case study, Z'-factor or SSMD may be too strict or too loose to be suitable for selection thresholds for hits even in experiments for which the metrics are designed. In addition, the thresholds of these measures have no intuitive meaning so we can not possibly set the correct thresholds for the experiments without referring similar experiments in the literature. However, due to the inherent variability of biological system, such as genetic and transcriptional change from unknown sources\cite{ben2018genetic}, even when the same experiment performed with the same cell type, the threshold provided in the literature may not be appropriate. 

In contrast, the proposed GSSMD measure is relatively sensitive to detect changes in biological assays. It is also robust against outliers and measurement noise, as described in simulation studies. Non-parametric characteristics may cause deviations in the estimates of the measurements, but we have shown that the appropriate number of samples and bin size can minimize this variability. In addition to this performance issue, GSSMD also showed the desired characteristics of the assay quality metric, i.e., interpretability. By definition, it is a measure of non-overlaping proportion of two distributions, so we can control lower bound of false positive rate of an assay through GSSMD threshold. Both performance and interpretability of GSSMD make it an attractive alternative to other assay quality and threshold selection methods. 

\section{Conclusion}
In this paper, we presented new assay quality metric called GSSMD. We showed that GSSMD may have advantages compare to the conventional quality control measures such as Z'-factor and SSMD based on simulation and real biological assay data set. We believe that the proposed method can be easily implemented in current HTS setup and medium throughput experimental setup. 

\section*{Acknowledgment}
We thank Alfonso Rodriguez-Molares (alfonso.r.molares@ntnu.no) for sharing a MATLAB implementation of GCNR, a similar task performed in ultra-sound image quality assessment. We also thank Yoonhyeok Lee for helpful discussion about applying GSSMD measure on the small scale scenario. 

\section*{Conflict of interest}
The authors reported no conflict of interest.




\bibliographystyle{IEEEtran}
\bibliography{REFERENCES}

\begin{thebibliography}{10}
\providecommand{\url}[1]{#1}
\csname url@samestyle\endcsname
\providecommand{\newblock}{\relax}
\providecommand{\bibinfo}[2]{#2}
\providecommand{\BIBentrySTDinterwordspacing}{\spaceskip=0pt\relax}
\providecommand{\BIBentryALTinterwordstretchfactor}{4}
\providecommand{\BIBentryALTinterwordspacing}{\spaceskip=\fontdimen2\font plus
\BIBentryALTinterwordstretchfactor\fontdimen3\font minus
  \fontdimen4\font\relax}
\providecommand{\BIBforeignlanguage}[2]{{%
\expandafter\ifx\csname l@#1\endcsname\relax
\typeout{** WARNING: IEEEtran.bst: No hyphenation pattern has been}%
\typeout{** loaded for the language `#1'. Using the pattern for}%
\typeout{** the default language instead.}%
\else
\language=\csname l@#1\endcsname
\fi
#2}}
\providecommand{\BIBdecl}{\relax}
\BIBdecl

\bibitem{broach1996high}
J.~R. Broach, J.~Thorner \emph{et~al.}, ``High-throughput screening for drug
  discovery,'' \emph{Nature}, vol. 384, no. 6604, pp. 14--16, 1996.

\bibitem{Malo2006StatisticalPI}
N.~Malo, J.~A. Hanley, S.~Cerquozzi, J.~Pelletier, and R.~Nadon, ``Statistical
  practice in high-throughput screening data analysis,'' \emph{Nature
  Biotechnology}, vol.~24, pp. 167--175, 2006.

\bibitem{Zhang1999ASS}
Q.~Zhang, Chung, and Oldenburg, ``A simple statistical parameter for use in
  evaluation and validation of high throughput screening assays.''
  \emph{Journal of biomolecular screening}, vol. 4 2, pp. 67--73, 1999.

\bibitem{Zhang2007APO}
X.~D. Zhang, ``A pair of new statistical parameters for quality control in rna
  interference high-throughput screening assays.'' \emph{Genomics}, vol. 89 4,
  pp. 552--61, 2007.

\bibitem{birmingham2009statistical}
A.~Birmingham, L.~M. Selfors, T.~Forster, D.~Wrobel, C.~J. Kennedy, E.~Shanks,
  J.~Santoyo-Lopez, D.~J. Dunican, A.~Long, D.~Kelleher \emph{et~al.},
  ``Statistical methods for analysis of high-throughput rna interference
  screens,'' \emph{Nature methods}, vol.~6, no.~8, p. 569, 2009.

\bibitem{rodriguez2019generalized}
A.~{Rodriguez-Molares}, O.~M.~H. {Rindal}, J.~{D'hooge}, S.~{Måsøy},
  A.~{Austeng}, M.~A.~L. {Bell}, and H.~{Torp}, ``The generalized
  contrast-to-noise ratio: a formal definition for lesion detectability,''
  \emph{IEEE Transactions on Ultrasonics, Ferroelectrics, and Frequency
  Control}, pp. 1--1, 2019.

\bibitem{cellucci2005statistical}
C.~J. Cellucci, A.~M. Albano, and P.~E. Rapp, ``Statistical validation of
  mutual information calculations: Comparison of alternative numerical
  algorithms,'' \emph{Physical Review E}, vol.~71, no.~6, p. 066208, 2005.

\bibitem{boutros2004genome}
M.~Boutros, A.~A. Kiger, S.~Armknecht, K.~Kerr, M.~Hild, B.~Koch, S.~A. Haas,
  R.~Paro, N.~Perrimon, H.~F.~A. Consortium \emph{et~al.}, ``Genome-wide rnai
  analysis of growth and viability in drosophila cells,'' \emph{Science}, vol.
  303, no. 5659, pp. 832--835, 2004.

\bibitem{furusawa2005ubiquity}
C.~Furusawa, T.~Suzuki, A.~Kashiwagi, T.~Yomo, and K.~Kaneko, ``Ubiquity of
  log-normal distributions in intra-cellular reaction dynamics,''
  \emph{Biophysics}, vol.~1, pp. 25--31, 2005.

\bibitem{boutros2006analysis}
M.~Boutros, L.~P. Br{\'a}s, and W.~Huber, ``Analysis of cell-based rnai
  screens,'' \emph{Genome biology}, vol.~7, no.~7, p. R66, 2006.

\bibitem{ran2013genome}
F.~A. Ran, P.~D. Hsu, J.~Wright, V.~Agarwala, D.~A. Scott, and F.~Zhang,
  ``Genome engineering using the crispr-cas9 system,'' \emph{Nature protocols},
  vol.~8, no.~11, p. 2281, 2013.

\bibitem{bray2016cell}
M.-A. Bray, S.~Singh, H.~Han, C.~T. Davis, B.~Borgeson, C.~Hartland,
  M.~Kost-Alimova, S.~M. Gustafsdottir, C.~C. Gibson, and A.~E. Carpenter,
  ``Cell painting, a high-content image-based assay for morphological profiling
  using multiplexed fluorescent dyes,'' \emph{Nature protocols}, vol.~11,
  no.~9, p. 1757, 2016.

\bibitem{ben2018genetic}
U.~Ben-David, B.~Siranosian, G.~Ha, H.~Tang, Y.~Oren, K.~Hinohara, C.~A.
  Strathdee, J.~Dempster, N.~J. Lyons, R.~Burns \emph{et~al.}, ``Genetic and
  transcriptional evolution alters cancer cell line drug response,''
  \emph{Nature}, vol. 560, no. 7718, p. 325, 2018.

\end{thebibliography}

\end{document}


\begin{figure}[htbp]
\centerline{\includegraphics[width = 15cm]{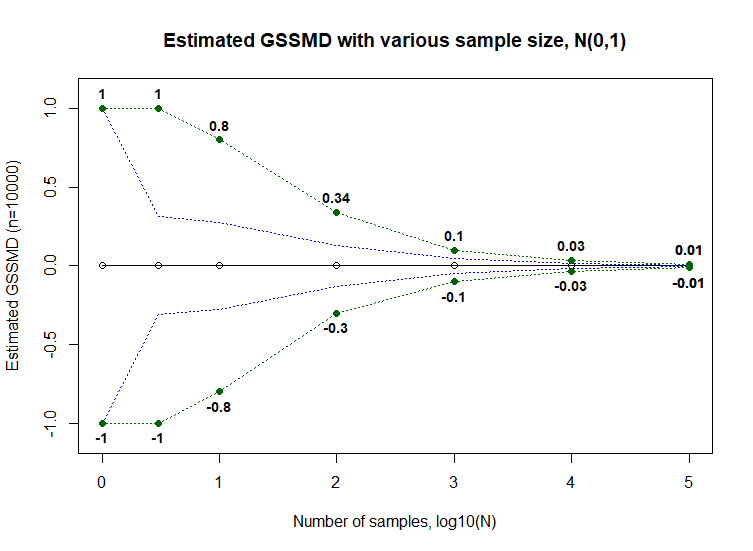}}
\caption{Effect of sample size on GSSMD when normal distributions overlap completely ($\mathcal{N}(0,1)$). The blue line represents the standard deviation and the green line represents the maximum and minimum of the estmated GSSMD. The number of independent trials is $n=10,000$}
\label{supp1}
\end{figure}

\begin{figure}[htbp]
\centerline{\includegraphics[width = 15cm]{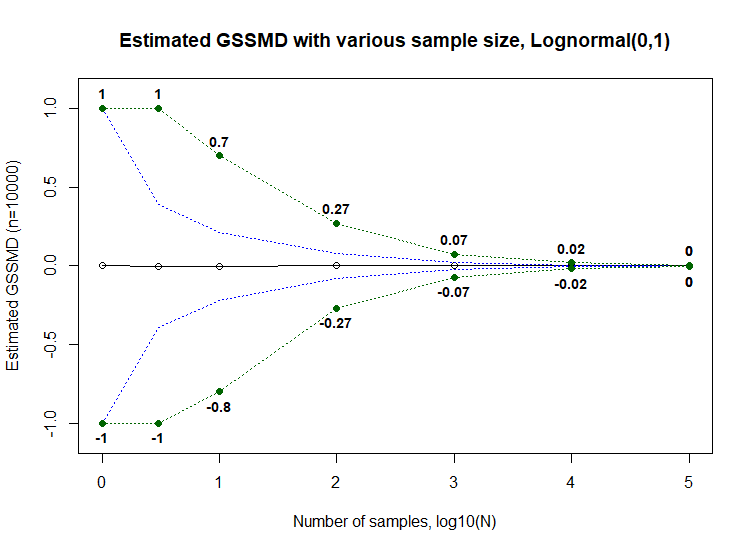}}
\caption{Effect of sample size on GSSMD when log-normal distributions overlap completely ($Log\mathcal{N}(0,1)$).  The blue line represents the standard deviation and the green line represents the maximum and minimum of the estmated GSSMD. The number of independent trials is $n=10,000$}
\label{supp2}
\end{figure}